\documentclass{emulateapj}
\shorttitle{Metalicity and migration}
\shortauthors{Dawson \& Murray-Clay}
\def\kep{\emph{Kepler\ }}
\def\kepdot{\emph{Kepler.\ }}
\def\fHJK{f_{\rm HJ, Kepler}}
\def\fHJRV{f_{\rm HJ, RV}}
\begin{document}
\slugcomment{Received 2013 February 6; accepted 2013 February 22; published 2013 April 2}
\title{Giant planets orbiting metal-rich stars show signatures of planet-planet interactions}
\author{Rebekah I. Dawson\altaffilmark{1}}
\author{Ruth A. Murray-Clay}
\affil{Harvard-Smithsonian Center for Astrophysics, 60 Garden St, MS-10, Cambridge, MA 02138}
\altaffiltext{1}{{\tt  rdawson@cfa.harvard.edu}}

\begin{abstract}
Gas giants orbiting interior to the ice line are thought to have been displaced from their formation locations by processes that remain debated. Here we uncover several new metallicity trends, which together may indicate that two competing mechanisms deliver close-in giant planets: gentle disk migration, operating in environments with a range of metallicities, and violent planet-planet gravitational interactions, primarily triggered in metal-rich systems in which multiple giant planets can form. First, we show with 99.1\% confidence that giant planets with semi-major axes between 0.1 and 1 AU orbiting metal-poor stars ([Fe/H]$<$0) are confined to lower eccentricities than those orbiting metal-rich stars. Second, we show with 93.3\% confidence that eccentric proto-hot Jupiters undergoing tidal circularization primarily orbit metal-rich stars. Finally, we show that only metal-rich stars host a pile-up of hot Jupiters, helping account for the lack of such a pile-up in the overall \kep sample. Migration caused by stellar perturbers (e.g. stellar Kozai) is unlikely to account for the trends. These trends further motivate follow-up theoretical work addressing which hot Jupiter migration theories can also produce the observed population of eccentric giant planets between 0.1 and 1 AU.
\end{abstract}

\section{Introduction}

Approximately 1\% of stars host hot Jupiters, ousted from their birthplaces to short-period orbits \citep{2012W} via mechanisms that remain debated. Proposed theories fall into two classes: smooth disk migration (e.g. \citealt{1980G}), and migration via gravitational perturbations, either by stars (e.g. stellar binary Kozai, \citealt{2003W}) or sibling planets (including planetary Kozai, e.g. \citealt{2011NF}; scattering, e.g. \citealt{1996R}; and secular chaos, e.g. \citealt{2011WL}). (See \citealt{2013D}, DMJ13 hereafter, for additional references.) We consider the latter class as also encompassing gravitational perturbations preceeded by disk migration (e.g. \citealt{2011G}).

Migration processes must not only produce hot Jupiters --- heavily studied, extensively observed gas giants orbiting within 0.1 AU of their host stars --- but also populate the region from 0.1 to 1 AU. This region is outside the reach of tidal damping forces exerted by the host star but interior to both the ice line and the observed pile-up of giant planets at 1 AU, one of which likely indicates where large, rocky cores can grow and accrete. We call this semi-major axis range the "Valley," because it roughly corresponds to the ``Period Valley" (e.g. \citealt{2003J}), the observed dip in the giant planet orbital period ($P$) distribution from roughly $10<P<100$ days. The Valley houses gas giants both on highly eccentric and nearly circular orbits. Gas disk migration is unlikely to excite large eccentricities (e.g. \citealt{2013DA}) whereas dynamical interactions are unlikely to produce a substantial population of circular orbits. Therefore this eccentricity distribution may point toward intermixing between two different migration mechanisms, one gentle and one violent. Another orbital feature --- the bimodal distribution of spin-orbit alignments among hot Jupiters --- is sometimes interpreted as evidence for two migration mechanisms \citep{2009F,2011M,2012N}. However, it may result from stellar torques on the proto-planetary disk \citep{2012B}, gravity waves that misalign the star's spin axis \citep{2012R}, or two regimes for tidal realignment \citep{2010W, 2012A}. Because tides are negligible in the Valley (except at the most extreme periastron, e.g. HD-80606-b, HD-17156-b), we can interpret trends more easily. Excited inclinations and eccentricities cannot have been erased by tidal damping.

If \emph{two} common mechanisms indeed deliver close-in giant planets, physical properties of the proto-planetary environment may determine which is triggered. A decade ago, \citet{2001S,2004S} discovered that giant planets more commonly orbit metal-rich stars, supporting the core accretion formation theory. Independent and follow-up studies confirmed this trend for giant planets (e.g. \citealt{2005F}, \citealt{2009S} \citealt{2010J}, \citealt{2011SS}, \citealt{2012MS}) but not small planets \citep{2007R,2012BL}. Neither the \citet{2004S} nor the \citet{2005F} samples exhibited correlations between stellar metallicity and planetary period or eccentricity, but now the radial-velocity (RV) sample has quadrupled. It is time to revisit the planet-metallicity correlation, but now to gain insight into the dynamical evolution of planetary systems \emph{following} planet formation.

Another motivation is the puzzlingly low occurrence rate of hot Jupiters in the \kep vs. RV sample \citep{2011Y,2012H,2012W}. \kep targets have systematically lower metallicities than RV targets. We will show that differences in the planetary period distribution --- not just the overall occurrence rate --- between metal-rich and metal-poor stars may account for the discrepancy.

We uncover new stellar metallicity trends in the eccentricities of giant Valley planets (\S2), eccentricities of giant planets tidally circularizing (\S3), and giant planet period distribution (\S4). These correlations point toward planet-planet interactions as one of two mechanisms for delivering close-in gas giants (\S5). 

\section{Eccentric Valley planets orbit metal-rich stars }
\begin{figure*}
\includegraphics{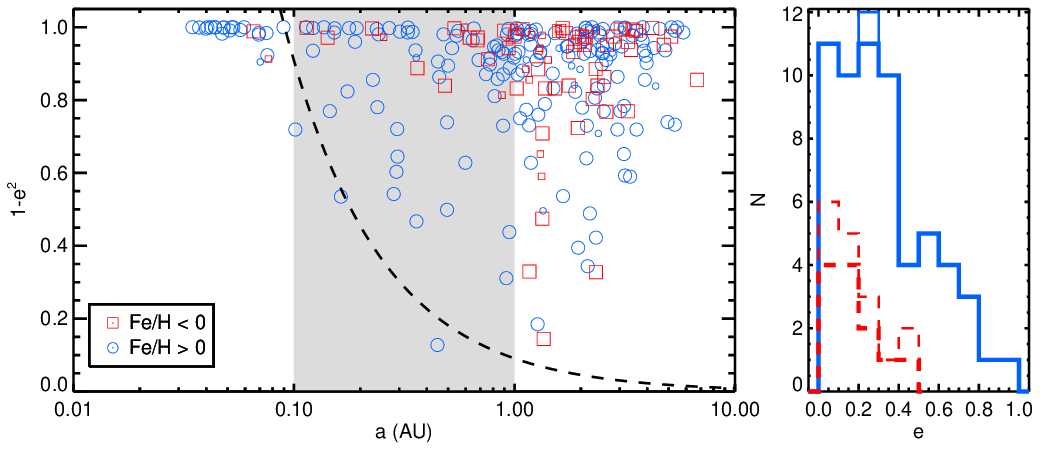}
\caption{ Left: Valley (gray region) giant planets orbiting metal-rich stars ([Fe/H]$\ge$0, blue circles) have a range of eccentricities; those orbiting metal-poor stars ([Fe/H]$<$0, red squares) are confined to low eccentricities. Small symbols represent stars with $\log g < 4$. For reference, above the dashed line (a tidal circularization track ending at 0.1 AU) planets are unlikely to experience significant tidal circularization. We plot the quantity $1-e^2$ to emphasize high-eccentricity planets. Right: Eccentricity distributions of Valley planets orbiting metal-rich (blue solid) and metal-poor (red dashed) stars. The bold distributions omit stars with $\log g < 4$.}
\end{figure*}

Valley gas giants are unlikely to have formed in situ \citep{2006R} and exhibit a range of eccentricities ($e$) (Figure 1). Here we consider giant planets discovered by radial-velocity surveys with $m\sin~i>0.1M_{\rm Jup}$, (queried from the Exoplanet Orbit Database\footnote{Five planets fulfilling our selection criteria have eccentricities fixed at 0 in the EOD fits. We perform Monte Carlo Markov Chain fits to the RVs of 14-And-b ($e=0.026_{-0.013}^{+0.016}$), HD-81688-b ($e=0.031_{-0.015}^{+0.020}$), and Xi-Aql-b ($e=0.26\pm0.04$) using \citet{2008S}'s data; adopt  \citet{2011J}'s $e=0.03(<0.28)$ for HD-96063-b; and remove HD-104067-b because the RVs are unavailable.} [EOD] on March 1st, 2013, \citealt{2011W}). We restrict the sample to FGK stars ($0.4 < M_\star < 1.4 M_\odot$).

Under the two migration mechanisms hypothesis, Valley planets on nearly circular orbits moved in smoothly through the gaseous proto-planetary disk, whereas those on eccentric orbits were displaced through multi-body interactions. In Figure 1, we emphasize planets with large eccentricities by plotting $1-e^2$. This quantity is related to the specific orbital angular momentum, $h=\sqrt{a(1-e^2)}$, an important parameter for dynamical interactions. This scale also minimizes eccentricity bias. For example, as a result of noise and eccentricity bias, a planet truly on a circular orbit could have a measured $e\sim0.1$. However, on this scale, $e=0.1$ would be nearly indistinguishable from $e=0$.

We divide the sample into planets orbiting metal-rich stars ([Fe/H]$\ge$0, blue circles) vs. metal-poor stars ([Fe/H]$<$0, red squares). Only the metal-rich stars host Valley planets with large eccentricities. The eccentricities of these 61 planets extend up to 0.93. In contrast, the 17 Valley planets orbiting metal-poor stars are confined to low eccentricities ($e\le0.43$). Overall, 28\% of Valley planets orbiting metal-rich stars have eccentricities exceeding that of the most eccentric one orbiting a metal-poor star. 

We assess the statistical significance of the low eccentricities of Valley planets orbiting metal-poor stars. We perform a Kolmogorov-Smirnov (K-S) test on the null hypothesis that the eccentricities of the metal-rich and metal-poor sample are drawn from the same distribution. We reject the null hypothesis with 95.1\% confidence. Using a test more sensitive to the tails of distributions, Anderson-Darling (A-D), we reject the null hypothesis with 96.9\% confidence. Finally the probability that the maximum eccentricity of the 17 planets is less than or equal to the observed $e = 0.43$ is the ratio of combinations:

\[ \frac{ \left( \begin{array}{c}
61~{\rm Valley}~e  \le 0.43 \\
17~{\rm Valley}~[Fe/H]<0 \end{array} \right)}
{\left( \begin{array}{c}
78~{\rm Valley} \\
17~{\rm Valley}~[Fe/H]<0 \end{array} \right)}  = 0.86\% \] 

The results are insensitive to the exact metallicity cut and significant at 95\% confidence or higher for any cut located between -0.15 and 0.03 dex. Therefore, with 99.14\% confidence, we reject the hypothesis that the confinement to low eccentricities of the planets orbiting metal-poor stars results from chance. Although the exact statistical significance is somewhat sensitive to the definition of the Valley, which defines the sample size, it is evident in Figure 1 that the trend occurs throughout the Valley, and the significance of the results is 95\% or higher for cuts from $0.6 < a < 1.16$ AU. The significance is 99.86\% without the stellar cuts and 97.8\% with an additional cut of $\log g > 4$ to remove evolved stars.

As suggested by \citet{2012J} in the context of the mutual inclinations of \kep multi-planet systems, one might expect a threshold metallicity to trigger instability. Decreasing planets' semi-major axes $(a)$ via gravitational perturbations requires interactions between at least two (and probably more) closely-spaced giant planets. It may be that only metal-rich proto-planetary environments can form such systems.\footnote{RV systems containing multiple known giant planets do appear to have systematically higher metallicities than those containing one, but the statistical significance is marginal. We note that planets may be scattered to distances beyond current RV detection or ejected, so systems with only one known giant planet perhaps originally had more.} In contrast, planets on circular orbits would have arrived via disk migration, which can occur regardless of metallicity. 

We note that beyond 1 AU, the metal-rich and metal-poor sample have similar eccentricity distributions. Planets with $a>1$ AU have not necessarily changed their semi-major axes: they may have formed where we observe them. These planets on eccentric orbits near their formation location may have exchanged angular momentum with another planet or star without requiring the abundance of closely-packed giant planets necessary to drastically alter $a$.

\section{Proto-hot Jupiters orbit metal-rich stars}

\begin{figure*}
\includegraphics{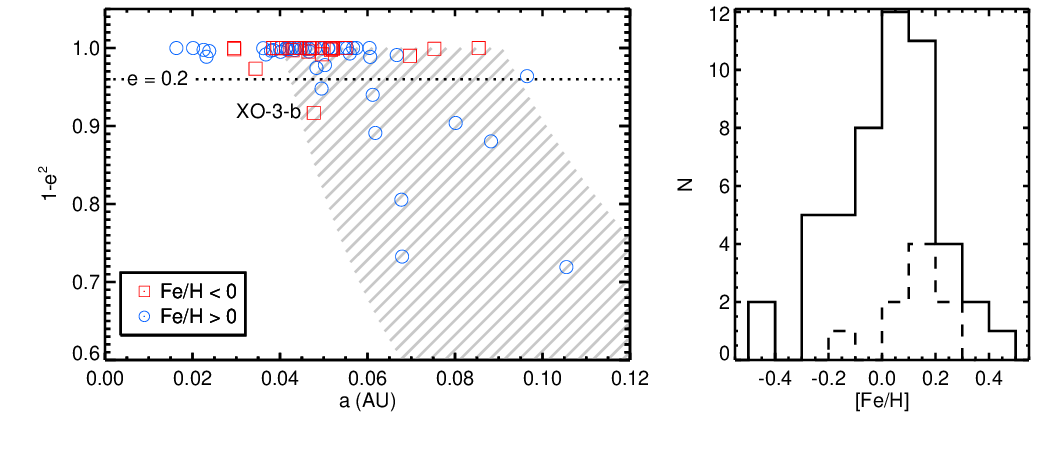}
\caption{Left: Giant planets discovered by non-\kep transit surveys, orbiting metal-rich (blue circles) and metal-poor (red squares) stars. The striped region encloses planets undergoing tidal circularization to $3<P_{\rm final}<10$ days. Planets below the dotted line have $e>0.2$, most of which orbit metal-rich stars. Right: Distribution of host star metallicities for planets in the striped region (left) with $e>$0.2 (dotted line) and $e<0.2$ (solid line).}
\end{figure*}

We turn to planets experiencing significant tidal dissipation, detected\footnote{Some planets have $e$ fixed at 0 in EOD fits. We remove those with poorly-constrained eccentricities: CoRoT-7-b, HAT-P-9-b, OGLE-TR-10-b, OGLE-TR-111-b, TrES-1-b, TrES-4-b, WASP-13-b, WASP-39-b, WASP-58-b, XO-1-b, XO-5-b. We include planets whose eccentricities are constrained to be small ($e<0.2$), by our fits (CoRoT-13-b, CoRoT-17-b, WASP-16-b) or the literature (CoRoT-7-b, HAT-P-1-b, HAT-P-4-b, HAT-P-8-b, HAT-P-12-b, HAT-P-27-b, HAT-P-39-b, OGLE-TR-211-b, KELT-2-Ab, WASP-7,  WASP-11-b, WASP-15-b, WASP-21-b, WASP-25-b, WASP-31-b, WASP-35-b, WASP-37-b, WASP-41-b, WASP-42-b, WASP-47-b, WASP-61-b, WASP-62-b, WASP-63-b, WASP-67-b). See the EOD for each planet's orbital reference.}  by non-\kep transit surveys (Figure 2) and followed up with RV measurements. We use the stellar and planetary cuts described in \S1  (except for XO-3-b, see below). \citet{2012S} and DMJ13 used this sample to calculate the abundance of moderately-eccentric proto-hot Jupiters. Advantageously for this sample, transit surveys are less inclined to target metal-rich stars, yielding planets orbiting metal-poor stars for comparison. To be consistent with \citet{2012S} and DMJ13 and to avoid eccentricity bias, we classify planets with $e>0.2$ as eccentric.

The striped region contains planets undergoing tidal circularization along tracks of constant angular momentum (see \citealt{2012S}, DMJ13) to final orbital periods $P_{\rm final}$ between 2.8 and 10 days. (The traditional boundary for hot Jupiters is 10 days, and 2.8 days is the limit above which we still see eccentric giant planets. Those with $P_{\rm final}<2.8$ days have much faster tidal circularization rates.) Most observed eccentric planets orbit metal-rich stars (blue circles). We suggest that only giant planets forming in metal-rich systems with multiple giant planets are likely to be scattered onto eccentric orbits that bring them close enough to the star to undergo tidal circularization (e.g. \citealt{2006F}).

The probability of randomly selecting eight planets orbiting stars with $[Fe/H] \ge 0$ and one planet (i.e. XO-3-b) orbiting a star with $[Fe/H] \ge -0.18$ is the ratio of combinations:
\[ \frac{ \left( \begin{array}{c}
38\\
8 \end{array} \right) \times 14 +  \left( \begin{array}{c}
38\\
9 \end{array} \right)}
{\left( \begin{array}{c}
59 \\
9 \end{array} \right)}  = 6.7\% \] 
where, among the 59 stars in the $P_{\rm final}$ range, 38 have $[Fe/H] \ge 0$ and 14 have $-0.18 \le [Fe/H] < 0$. XO-3 has $M_\star = 1.41 M_\odot$, just above our stellar mass cut; the high mass of the star (corresponding to a more massive disk and more metals to form giant planets) may account for the presence of a proto-hot Jupiter despite the star's low metallicity. Without this star, the statistical significance is 98.3\%. We also perform a K-S (A-D) test, rejecting with 95.5\% (92.1\%) confidence the null hypothesis that the host star metallicities of planets in the striped region with $e>$0.2 are drawn from the same distribution as those with $e<$ 0.2.

\section{The short-period pile-up is a feature of metal-rich stars}

\begin{figure}
\includegraphics{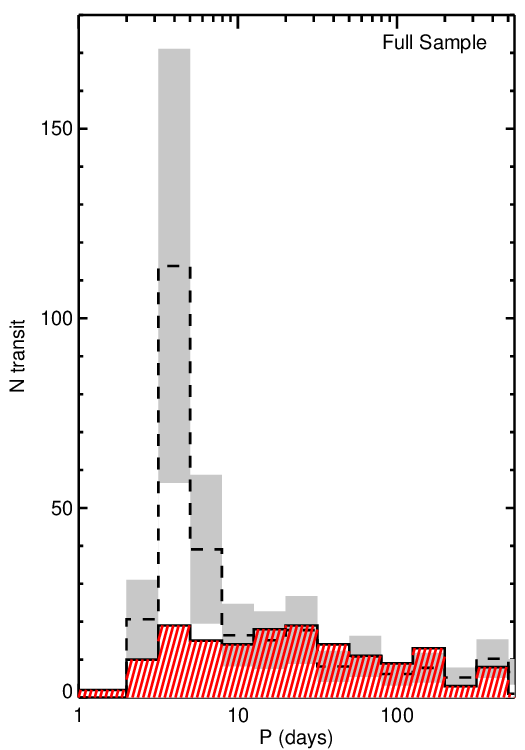}
\caption{Red striped: number of transiting giant planets detected by \kepdot Black dashed: expected number based on the RV-discovered (i.e. excluding planets discovered by transit surveys) sample$^{\ref{note:sample}}$. The gray error bars are from uncertainties in $C_{\rm norm}$, not the Poisson uncertainties of each individual bin. The two distributions are consistent at long periods, but the \kep sample lacks a short period pile-up.}
\end{figure}

\begin{figure}
\includegraphics{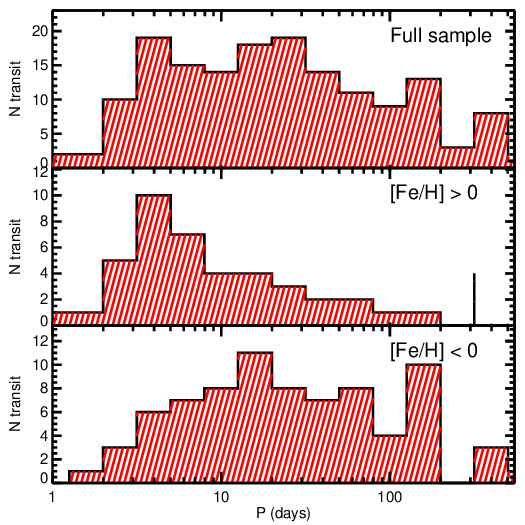}
\caption{Number of transiting giant planets observed by \kep without a stellar metallically cut (top),  with [Fe/H]$\ge$0 (middle), and with [Fe/H]$<$0 (bottom). In the metal-rich sample (middle), we recover the shape of the short-period pile-up seen in the RV sample (black-dashed line, Figure 3). In contrast, the metal-poor sample (bottom) is depleted in short-period giants.}
\end{figure}

\begin{figure*}
\includegraphics{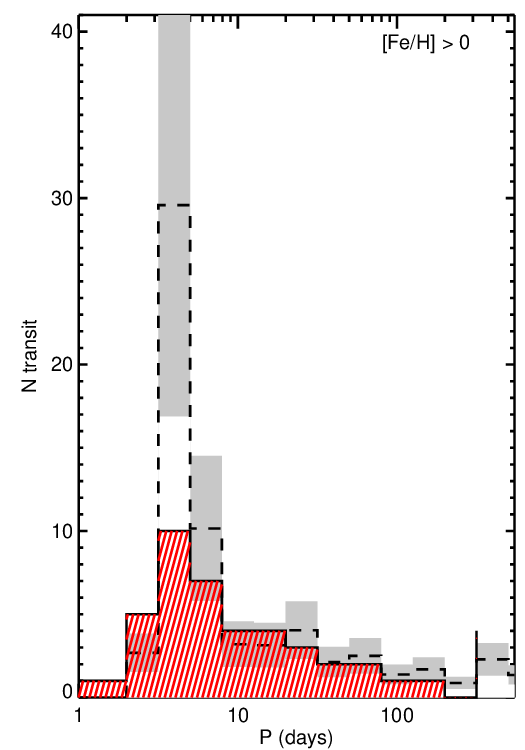}
\includegraphics{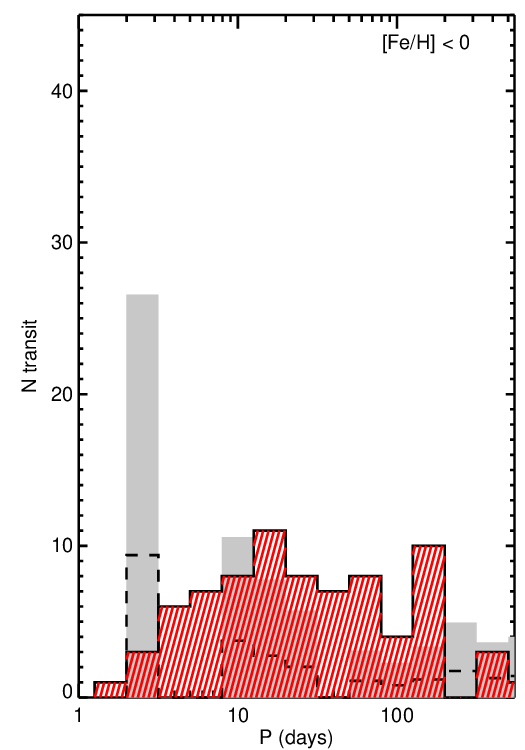}
\caption{Same as Figure 3 but for metal-rich (left) and metal-poor (right) subsamples. Left: Metal-rich \kep sample (red striped) exhibits a short-period pile-up, but falls below RV expectations in the 3-5 day bin. Right: Metal-poor \kep sample is not inconsistent with the metal-poor RV sample, but the latter is difficult to characterize due to small numbers.}
\end{figure*}

\citet{2012H} found a surprisingly low \kep hot Jupiter occurrence rate ($\fHJK$) --- the expected number of giant planets per star with $P<10$ days --- compared to RV surveys ($\fHJRV$), a trend confirmed by \citet{2012W} and \citet{2013F}; all suggested that the systematically lower metallicities of \kep host stars may contribute to the discrepancy. In Figure 3, we compare the period distribution of transiting giant planet candidates detected by the \kep survey (\citealt{2013BB}; see also \citealt{2011BK} and \citealt{2013BR}) --- applying a radius cut of 8$<R_{\rm planet}<20R_{\rm earth}$ --- to that expected from the RV sample,\footnote{\label{note:sample} The RV sample is not uniform; we plot it for qualitative comparison. The expected distribution derived from the period distribution reported by \citet{2008C} appears similar. We therefore interpret the short-period pile-up as real, not due to preferential detection. For the quantitative calculations in this section, we use the uniform \citet{2005F} sample.} using a normalization constant $C_{\rm norm}$ (defined below). The RV sample includes only planets discovered by RV surveys, not transit surveys. For both samples, we follow DMJ13 and impose cuts of stellar temperature $4500 < T < 6500 K$ and surface gravity $\log g > 4$ to restrict the sample to well-characterized Kepler host stars \citep{2011B}. The two distributions appear consistent beyond 10 days but differ strikingly at short orbital periods: the \kep period distribution lacks a short-period pile-up (in fact, the absolute \kep giant planet occurrence declines toward short orbital periods, as modeled by \citealt{2011Y} and \citealt{2012H}).

Although \kep Input Catalog (KIC) metallicity estimates are known to be uncertain \citep{2011B}, we can roughly divide the \kep sample into metal-rich ([Fe/H]$\ge$0) and metal-poor ([Fe/H]$<$0). In Figure 4, we compare the period distributions for \kep giant planets orbiting metal-rich vs. metal-poor stars. When we limit the sample to [Fe/H]$\ge0$ (row 2), we recover the missing short-period pile-up, which the metal-poor sample (row 3) lacks. Performing a K-S test, we reject with 99.95\% confidence the hypothesis that the metal-rich sample and metal-poor sample are drawn from the same distribution. The results are insensitive to the exact metallicity cut.

We compare the \kep metal-rich(poor) sample to the RV metal-rich(poor) sample in Figure 5. In Figures 3 and 5, we compare the observed number of transiting \kep giant planets (red striped) to the number expected (black dashed) based on the RV sample, $$N_{\rm RV, trans}=C_{\rm norm}N_{\rm RV}{\rm prob}_{\rm trans},$$ where $N_{\rm RV}$ is the observed number of RV planets per bin and ${\rm prob}_{\rm trans}(P)$ is the transit probability. We set the normalization constant, $C_{\rm norm}$, using the values (computed below) of $\fHJK$ and $\fHJRV$:
$$C_{\rm norm}=\frac{\fHJRV}{\fHJK}\frac{\sum_{P = 0}^{10 \rm days}N_{ \rm trans,Kep}(P)/{\rm prob}_{\rm trans}(P)}{\sum_{P = 0}^{10 \rm days}N_{\rm RV} (P) }.$$ 
\noindent Each error bar is due to the uncertainty in $\fHJRV/\fHJK$. To compute $\fHJK$, we follow \citet{2012H}, using our own stellar and planetary cuts and the latest sample of \kep candidates \citep{2013BB}. The Barbara A. Mikulski Archive for Space Telescopes (MAST) supplied the stellar parameters and the NExSci Exoplanet Archive the transit shape parameters (duration, depth, $a/R_\star$, $R_{\rm planet}/R_\star$). We obtain\footnote{We estimate the occurrence rates and uncertainties based on the Poisson likelihood and a Jeffrey's prior, following DMJ13.} $\fHJK=0.38^{+0.08}_{-0.07}\%$ for giant planets with $P<$10 days (consistent with \citealt{2012H} and \citealt{2013F}), 1.08$^{+0.33}_{-0.27}\%$ for the metal-rich sample, and 0.25$^{+0.08}_{-0.06}\%$ for the metal-poor sample. To compute $\fHJRV$, we use the stellar and planetary sample from the iconic planet-metallicity correlation \citep{2005F} and associated stellar parameters \citep{2005V}, the last RV target list to be publicly released. We obtain $\fHJRV=1.03^{+0.34}_{-0.32}$\% for giant planets with $P<$10 days (in agreement with \citealt{2012W}), 1.74$^{+0.67}_{-0.54}$\% for those orbiting stars with [Fe/H]$\ge$0, and 0.07$_{-0.06}^{+0.23}\%$ for  [Fe/H]$<$0. With no metallicity cut, $\fHJK$ is inconsistent with $\fHJRV$ at the $2.0 \sigma$ level.

In the metal-rich comparison (Figure 5, left), we see greater consistency between the \kep and RV distribution than in the full sample (Figure 3). The metal-rich \kep sample exhibits a short-period pile-up; the discrepancy between $\fHJK$ vs. $\fHJRV$ is now only $1.0\sigma$, with the greatest discrepancy in the 3-5 day bin. This improvement motivates a detailed follow-up analysis, including a more precise estimate of $\fHJRV$ using the latest RV target lists. If follow-up studies find a significant discrepancy between the metal-rich \kep and radial velocity samples, it could be due to the KIC metallicity estimates. Using spectroscopic metallicity measurements by \citet{2012BL}, we find that high metallicities do correspond linearly to high spectroscopic metallicities (with a scatter of about 0.2 dex about a best-fit line with slope 0.3), but the spectroscopic metallicities have a systematic offset corresponding to 0.1 dex at KIC [Fe/H] = 0, consistent with the discussion by \citet{2011B}. However, we attribute the systematic offset to the fact that stars targeted for spectroscopic follow-up are bright, main-sequence stars in our solar neighborhood and thus have systematically higher metallicities; in contrast, the KIC metallicities were computed assuming a low-metallicity prior, due to the \kep targets being above the galactic plane. The planetary radius cut may also contribute to the discrepancy.  The $8 R_{\rm earth}$ cut for the \kep sample corresponds to the RV cut of $m\sin$$i=0.1M_{\rm Jup}$ for a planet made of pure hydrogen at a low effective temperature (e.g. \citealt{2007S}). However, close-in, low-mass planets may be inflated to $>8R_{\rm earth}$ and may have a different period distribution, contaminating the sample. In the metal-poor comparison (Figure 5, right), the \kep and RV distributions do not appear inconsistent, but it is difficult to judge given the very small sample of RV-detected planets orbiting metal-poor stars.

\section{Conclusion}

We found three ways in which the properties of hot Jupiters and Valley giants depend on host star metallicity:

\begin{enumerate}
\item Gas giants with $a<1$AU orbiting metal-rich stars have a range of eccentricities, whereas those orbiting metal-poor stars are restricted to lower eccentricities.
\item Metal-rich stars host most eccentric proto-hot Jupiters undergoing tidal circularization.
\item The pile-up of short-period giant planets, missing in the \kep sample, is a feature of metal-rich stars and is largely recovered for giants orbiting metal-rich \kep host stars.
\end{enumerate}

Hot Jupiters and Valley giants are both thought to have been displaced from their birthplaces. Therefore these metallicity trends can be understood if smooth disk migration and planet-planet scattering both contribute to the early evolution of systems of giant planets. We expect disk migration could occur in any system, but only systems packed with giant planets -- which most easily form around metal-rich stars -- can scatter giant planets inward to large eccentricities (Trend 1). Some of these tides shrink and circularize (Trend 2), creating a pile-up of short-period giants (Trend 3). Moreover, these trends support planet-planet interactions (e.g. scattering, secular chaos, or Kozai) as the dynamical migration mechanism for delivering close-in giant planets, rather than stellar Kozai. This is consistent with previous work by DMJ13 arguing that stellar Kozai does not produce most hot Jupiters, based on the lack of super-eccentric proto-hot Jupiters. We would not expect planet-planet scattering to typically result in nearby companions to hot Jupiters, which have been ruled out in the \kep sample by \citet{2012SR}. (See also \citealt{2011L}.)

One possible challenge for our interpretation is the lack of apparent correlation between spin-orbit misalignment and metallicity. However, spin-orbit misalignments are not necessary caused by dynamical perturbations, and their interpretation is complicated because measurements have primarily been performed for close-in planets subject to tidal realignment. We recommend spin-orbit alignment measurements, via spectroscopy \citep{1924M,1924R,2000Q} or photometry \citep{2011N,2011S}, of \kep candidates in the Valley, which are typically too distant to be tidally realigned.

To support or rule-out the interpretation that these metallicity trends are signatures of planet-planet interactions, we further recommend: 1) theoretical assessments of whether planet-planet interaction mechanisms designed to account for hot Jupiters can simultaneously produce the observed population of eccentric Valley planets, and 2) more sophisticated assessments of the trends we report here, using the target lists of recent RV surveys and, as undertaken by \citet{2013F}, a careful treatment of \kep false-positives and detection thresholds.
\acknowledgments
The NSF-GRFP (DGE-1144152) supports R.I.D. We thank the referee for the helpful, timely report. Our gratitude to John Johnson for many illuminating discussions about the three-day pile-up, the period distribution of giant planets, observational approaches to distinguishing the origins of hot Jupiters, valuable insights on previous collaborations connected to this investigation, and extensive comments. We thank Daniel Fabrycky for many helpful comments, Courtney Dressing for occurrence rate insights, and Subo Dong, Zachory Berta, David Charbonneau, Sean Andrews, Matthew Holman, Jason Wright, and Kevin Schlaufman for useful discussions. We thank B. Scott Gaudi and Andrew Gould for helpful comments and corrections, including altering us about XO-3-b.

We used the Exoplanet Orbit Database and Exoplanet Data Explorer at exoplanets.org, and the NASA Exoplanet Archive, operated by Caltech, under contract with the NASA Exoplanet Exploration Program.  We include data collected by the \kep mission, funded by the NASA Science Mission directorate, and are grateful to the \kep Team for long and extensive efforts in producing this rich dataset. Some data were obtained from MAST. The Association of Universities for Research in Astronomy, Inc. operates STScI under NASA contract NAS5-26555. The NASA Office of Space Science supports MAST for non-HST data via grants (including NNX09AF08G) and contracts. 

Note added in proof: We thank readers for alerting us to references we missed. \citet{1997G} first pointed out the planet-metallicity correlation, \citet{2006G} that RV surveys are biased toward higher metallicities, and, in a submitted conference proceedings, \citet{2012T} a correlation between eccentricity and metallicity for close-in giant planets.

\bibliographystyle{apj}

\end{document}